\documentclass[runningheads]{llncs}
\usepackage[T1]{fontenc}
\usepackage{graphicx}
\usepackage{cite}
\usepackage{microtype}
\usepackage{pdfpages}

\begin{document}

\title{Federated Semantic Knowledge Graphs for Laboratory Workflows: A Structured Expert Elicitation Methodology Demonstrated Through Bioanalytical Workflow Twins}

\titlerunning{Federated SKGs for Laboratory Workflows}

\author{Luis F. Schachner\inst{1}\orcidID{0000-0001-6157-0937} \and
Vinith Thamizhazhagan\inst{2} \and
Sara Tanenbaum\inst{2} \and
John C. Tran\inst{1} \and
Pamela P. F. Chan\inst{1} \and
Mandy Kwong\inst{1} \and
Andy Chang\inst{1} \and
Maureen Beresini\inst{1} \and
Margaret Porter Scott\inst{1}}

\authorrunning{L. F. Schachner et al.}

\institute{Biochemical and Cellular Pharmacology Department, Genentech, Inc., South San Francisco, CA, USA \\
\email{schachl1@gene.com} \and
Computational Sciences Center of Excellence, Genentech, Inc., South San Francisco, CA, USA}

\maketitle

\begin{abstract}
Laboratory workflows in pharmaceutical and biomedical research encode substantial tacit knowledge — expert judgment about failure conditions, decision branching logic, and contextual dependencies — that remains inaccessible to protocol documents, sensor streams, and existing biomedical ontologies. We present a repeatable structured expert elicitation methodology and federated Semantic Knowledge Graph (SKG) architecture for capturing and querying this knowledge, demonstrated through deployment at the Biochemical and Cellular Pharmacology Department of Genentech, Inc. Knowledge is elicited via the Protocol Intelligence Co-pilot, a purpose-built AI interview agent that applies structured elicitation lenses to surface tacit procedural knowledge with expert-assigned confidence scores, producing graph representations across three tiers: program-level decision milestones, assay protocol knowledge, and physical execution infrastructure. Separately constructed subgraphs — exemplified by immunoassay (ELISA), quantitative mass spectrometry (LC-MS/PRM), and laboratory automation — are aligned through a shared upper ontology and queried as a single federated graph. Evaluation demonstrates seven query types structurally unavailable from any individual data source, including a cross-subgraph traversal that identifies automation-masked silent failures — conditions where execution logs report success while scientific validity is actively compromised. Most critically, the \texttt{MASKED\_BY} graph relationship encodes a class of laboratory risk invisible to current informatics platforms — the structural gap that prevents existing systems from reasoning about scientific validity. This architecture provides the semantic world model that AI laboratory agents currently lack: a queryable representation of where workflows fail silently, where human judgment is irreplaceable, and which execution assets mask rather than detect failure.

\end{abstract}


\section{Introduction}
Laboratory workflows in pharmaceutical and biomedical research encode knowledge that resists all existing capture approaches. A bioanalytical scientist running an immunoassay does not simply execute a sequence of physical steps --- she applies accumulated expert judgment about failure modes that manifest silently, decision logic that branches conditionally on results that look normal, and contextual dependencies that no protocol document has ever articulated. When she leaves the organization, this knowledge leaves with her. When she trains a junior colleague, the knowledge degrades. When an AI agent is asked to interpret an anomalous result, it operates without the scientific world model required to reason the way she does. The fundamental problem is not a lack of data --- modern laboratory systems produce enormous volumes of it. The problem is that none of the existing systems capture the semantic knowledge underneath: tacit procedural judgment, with its conditional structure, its confidence gradations, and its failure genealogy.

This gap is most sharply illustrated by automation-induced silent failures --- conditions where the laboratory execution system reports a successful run while the underlying scientific validity of the assay is compromised. A plate washer completing six aspiration-dispense cycles logs a successful wash operation regardless of whether residual buffer remains in the wells; if it does, the standard curve is contaminated and the assay fails silently. The data system receives clean optical density values, passes them downstream, and the erroneous result is incorporated into a program decision. No existing informatics platform --- ELN, instrument log, or LIMS --- encodes the knowledge required to recognize this failure.

We present a federated Semantic Knowledge Graph (SKG) architecture and a repeatable structured expert elicitation methodology for capturing this knowledge and making it queryable, deployed at the Biochemical and Cellular Pharmacology (BCP) Department at Genentech. Three independently constructed subgraphs --- immunoassay (ELISA), quantitative mass spectrometry (LC-MS/PRM), and laboratory automation infrastructure --- are aligned through a shared upper ontology and deployed in a single Neo4j AuraDB instance~\cite{ref29}. To elicit tacit knowledge from domain experts, we built the Protocol Intelligence Co-pilot --- an AI interview agent that applies structured elicitation lenses to surface tacit procedural knowledge, assigns confidence scores, and converts session outputs into MERGE-idempotent Cypher via a downstream annotator agent.

We evaluate the federation across six capability classes implemented as seven queries (Q1--Q7, \S 5), each demonstrating query types structurally unavailable from protocol documents, instrument sensors, or existing biomedical ontologies. Pipeline extraction is deterministic: independent re-runs on the same transcript yield identical failure mode identification (within-agent FM F1 = 1.0, zero variance, \S 5.3); cross-agent agreement is bounded by elicitation depth, reaching FM F1 = 1.0 on clean LC-MS transcripts and FM F1 = 0.43 on ELISA cross-agent comparison --- a gap explained by multi-turn conversational probing versus single-shot extraction (\S 5.3).

\section{Related Work}
Our work sits at the intersection of five literature bodies: semantic digital twins with knowledge graphs, self-driving laboratories and lab automation, structured expert elicitation, biomedical workflow knowledge representation, and the formal treatment of automation-induced silent failures. ISWC In-Use Track deployments of knowledge graphs in industrial and scientific settings provide the direct venue context \cite{ref22, ref26, ref34}.

\subsection{Semantic Digital Twins with Knowledge Graphs}
Knowledge graphs increasingly back digital twins across manufacturing, smart infrastructure, and industrial cyber-physical systems \cite{ref18, ref21, ref25, ref27, ref30, ref32, ref35, ref36}. These systems universally assume full observability via sensors. Our laboratory workflows violate this assumption: the scientific validity of a result is not observable from execution data. This observability gap is the representational problem that motivates our federated SKG.

The closest methodology to ours in spirit is D'Amico et al.~\cite{ref7}, who propose a five-step framework for Cognitive Digital Twins, establishing the concept of expert knowledge encoded into a formal graph for reasoning and decision support. Jungmann and Lazarova-Molnar~\cite{ref19} independently identify the same integration gap --- data-driven DTs lack systematic expert knowledge incorporation --- but leave the elicitation mechanism and uncertainty representation as open problems. Our work addresses both.

The BCP upper ontology draws on the Allotrope Foundation Ontology (AFO)~\cite{ref2} and the Ontology for Biomedical Investigations (OBI)~\cite{ref24}. Both are grounded in the Basic Formal Ontology (BFO) --- a realist ontology bounded by design to concretized objects and processes, not epistemic states or potential failure conditions. This realist boundary explains precisely why no BFO-grounded standard addresses the epistemic layer this work targets. BCP's \texttt{FailureMode}, \texttt{DecisionPoint}, \texttt{MASKED\_BY}, and confidence scoring infrastructure are additive extensions into that vocabulary space; all existing analytical data standard terms remain intact.

\subsection{Self-Driving Laboratories and Lab Automation}
Self-driving laboratories (SDLs) use KGs for autonomous orchestration \cite{ref1, ref8, ref9, ref38}. High-profile implementations include The World Avatar~\cite{ref4} --- a distributed KG for cross-site autonomous synthesis --- and MATTERIX~\cite{ref9}, a GPU-accelerated simulation framework for robotics-assisted chemistry digital twins. In pharmaceutical bioanalysis specifically, Thieme et al.~\cite{ref37} describe deep Opentrons integration under FAIR principles; PyLabRobot~\cite{ref40} provides cross-platform liquid-handler interfaces that represent the infrastructure heterogeneity our \texttt{UseCase} vocabulary bridges.

SDLs are effective where the experimental objective --- yield, purity, conversion --- is directly measurable and the optimization function can be specified. For bioanalytical science, neither condition holds: the objective involves matrix interference characterization, silent failure risk assessment, and cross-study comparability whose validity is not recoverable from execution logs.

\subsection{Expert Elicitation, Biomedical KGs, and Workflow Knowledge}
Structured expert elicitation (SEE) is an established methodology for formally capturing uncertain quantities from domain experts \cite{ref6, ref10, ref13, ref14, ref33}, but traditionally targets probability distributions for parameters, not procedural workflow structures. Our work is, to our knowledge, the first application of SHELF-grounded~\cite{ref13} elicitation to encode laboratory procedural knowledge into a property graph with confidence scoring. The closest prior art is knowledge elicitation for medical laboratory diagnostic expert systems~\cite{ref23}, which targets diagnostic rule extraction rather than procedural workflow capture --- a distinct problem. Zhang et al.~\cite{ref42} introduced conversational ontology requirements elicitation via LLMs; our Protocol Intelligence Co-pilot extends this direction into operational workflow capture, targeting failure genealogy and decision logic rather than ontology schema. Confidence scoring uses linguistic approximation grounded in Lakoff~\cite{ref20} and Zadeh~\cite{ref41}.

Large-scale biomedical KGs from literature mining~\cite{ref12, ref43} and representation learning~\cite{ref28} establish KG infrastructure for biomedical entity relationships. Ours is a different kind: a procedural workflow graph encoding tacit expert knowledge about how assays are run, where they fail, and what judgment is required at each decision point. No published KG for bioanalytical assay workflows was identified in our literature search. The closest published work is Schr\"{o}der et al.~\cite{ref31}, who demonstrate structure-based knowledge acquisition from ELN protocols for provenance documentation; our contribution extends this into tacit procedural knowledge, automation-masked scientific validity, and explicit confidence scoring.

\subsection{Automation-Induced Silent Failures and the Observability Gap}
Avizienis et al.~\cite{ref3} define silent failures as components that fail without generating any error signal --- the formal backbone for our \texttt{MASKED\_BY} relationship: an automation asset that logs successful operation while scientific validity is compromised produces exactly this behavior.

The closest formal ontology work is MALFO~\cite{ref5}, a BFO-compatible ontology of malfunction-related occurrents (FOIS 2024), which formalizes a precise taxonomy of engineering failures aligned with BCP's internal failure taxonomy. MALFO does not address the automation reporting layer --- by design --- the case where an instrument's success log actively masks an underlying failure condition. The IMDRF adverse event terminology~\cite{ref17} recognizes device output errors from a regulatory reporting perspective but does not address the automation-masking pattern in experimental workflows. Regulatory guidance frameworks recognize that automation success does not guarantee scientific validity but provide no encoding mechanism~\cite{ref11, ref15, ref16}. The \texttt{MASKED\_BY} relationship and \texttt{silent\_failure\_risk} property are our encoding of exactly that divergence.

\section{Federated Architecture and Upper Ontology}
The Biochemical and Cellular Pharmacology Department (BCP) at Genentech, Inc. is the evidentiary backbone of the drug discovery pipeline: program milestones from candidate nomination through IND are gated on specific assay outputs that BCP designs, executes, and interprets. We use ``Semantic Digital Twin'' to denote a knowledge-anchored, query-first semantic model of a real system --- static by design in this deployment, not a live-synchronized physical replica; transition to real-time agent-runtime querying is the primary future work direction (\S 7). The BCP SDT is a federated SKG deployed in Neo4j AuraDB~\cite{ref29} whose architecture mirrors BCP's decision structure directly across three tiers, exemplified by three independently constructed subgraphs --- immunoassay (ELISA), quantitative mass spectrometry (LC-MS/PRM), and laboratory automation infrastructure --- coexisting in a single property graph database, aligned through a shared upper ontology.

\subsection{Three-Tier Knowledge Hierarchy}
\begin{figure}
    \centering
    \includegraphics[width=1\textwidth,keepaspectratio]{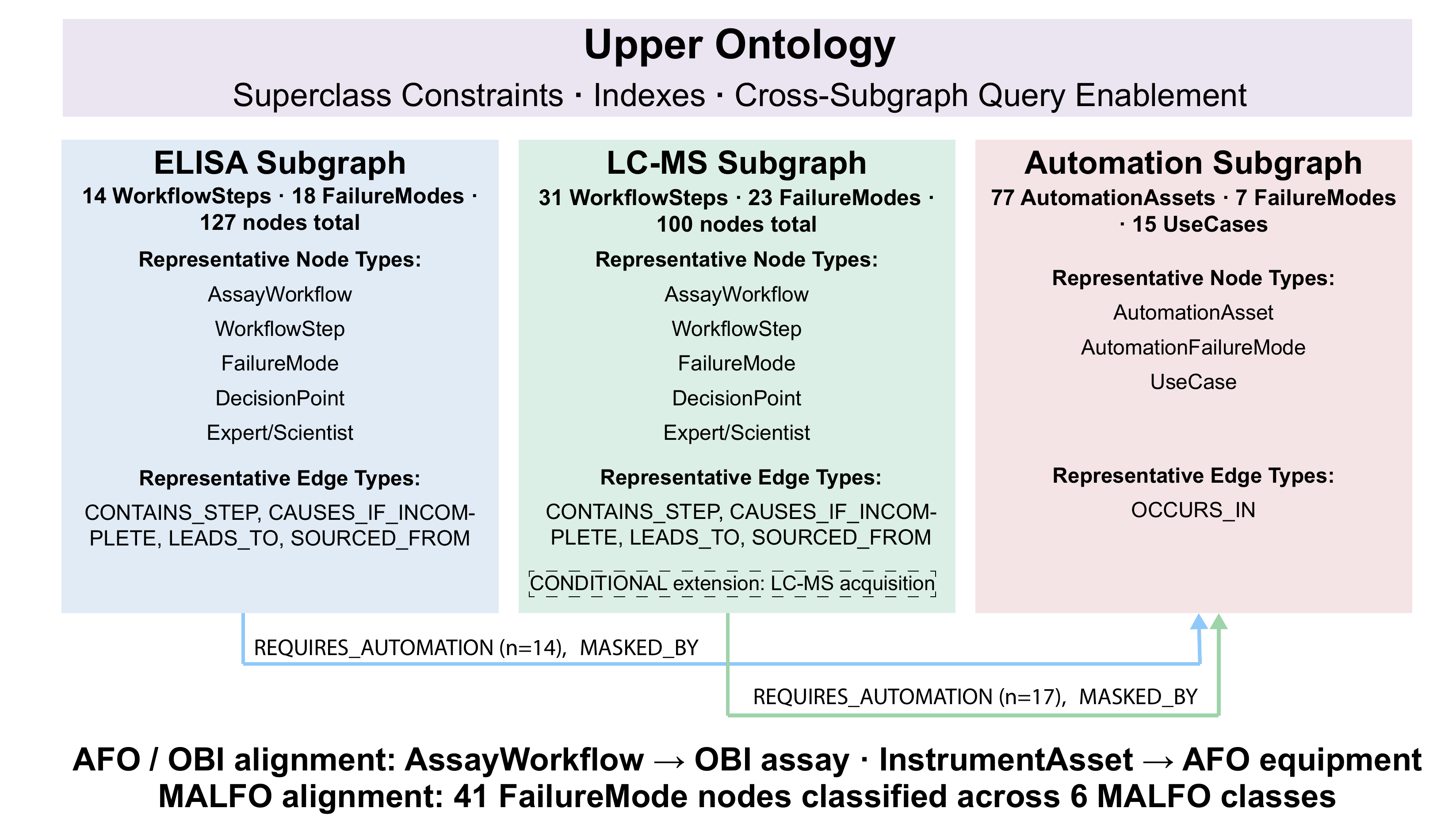}
    \caption{BCP Semantic Digital Twin --- federated architecture. Three subgraphs are linked through cross-subgraph edges; the shared upper ontology provides alignment anchors at the AssayWorkflow and AutomationAsset layers. The CONDITIONAL block marks the LC-MS modality-specific extension.}
    \label{fig1}
\end{figure}

The architecture organizes knowledge across three tiers that directly mirror the decision-making structure of a pharmaceutical research organization (Fig.~\ref{fig1}):

\begin{itemize}
    \item \textbf{Tier 1 --- Program Decision Layer.} Maps how assay outputs feed into program-level decisions. The defining edge type is \texttt{SOURCED\_FROM}, where each \texttt{EvidentiaryInput} node specifies the required assay output, quality threshold, and decision consequence if unmet. A cross-tier traversal from Tier 1 through Tier 3 therefore exposes whether the execution infrastructure supporting a specific program milestone carries undetected silent failure risk --- a query unresolvable from any single-tier data source.
    \item \textbf{Tier 2 --- Assay Protocol Layer.} The core scientific knowledge tier: workflow steps, decision logic, and FailureMode risks.
    \item \textbf{Tier 3 --- Execution Infrastructure Layer.} Models the physical automation environment and instrument logs (ErrorSignature), enabling representation of the gap between sensor scope and scientific consequence.
\end{itemize}

The three tiers are linked by a connection layer of cross-tier edge types described in \S 3.4.

\subsection{Upper Ontology Design and AFO Alignment}
The BCP upper ontology serves two purposes: enabling cross-subgraph query through shared superclasses, and grounding vocabulary in established laboratory analytical science terminology.

\subsubsection{Shared terminological layer (TBox).} The upper ontology defines shared superclasses (TBox); domain-specific subgraphs contain contextual instantiated data (ABox). Observable entities (processes, instruments) are grounded in AFO~\cite{ref2} and OBI~\cite{ref24}; we extend these standards to capture epistemic states --- conditional judgment, confidence gradations, automation-masked validity --- that lie outside BFO's intentional realist scope.

\subsubsection{Cross-subgraph query alignment.} A query against the \texttt{FailureMode} superclass traverses both ELISA and LC-MS subgraphs without modification; domain-specific properties coexist on the same node without conflicting with universally defined properties.

Table 1 (supplemental) documents the domain vocabulary basis for each BCP upper ontology superclass. The coverage pattern is informative: established industry standards provide vocabulary for the observable layer --- processes, instruments, analytes, measurement results --- and have no vocabulary for the epistemic layer --- failure genealogy, conditional expert judgment, tacit knowledge confidence, and automation-masked validity. This boundary is our contribution's entry point.

A note on ontological scope: AFO and OBI are grounded in BFO, a deliberately realist ontology that models concretized objects and processes but not epistemic states or potential failure conditions --- the representational gap this work addresses. The BCP schema is an application-level property graph vocabulary, not a formal ontology in the W3C/OBO sense (no minted IRIs, Aristotelian definitions, or OWL axioms); formal OWL serialization with DOLCE or GFO alignment --- more appropriate foundations for epistemic and dispositional concepts than BFO --- is scoped as future work.

\subsection{Node Type Schema}
\begin{figure}[htbp]
\centering
\includegraphics[width=\textwidth]{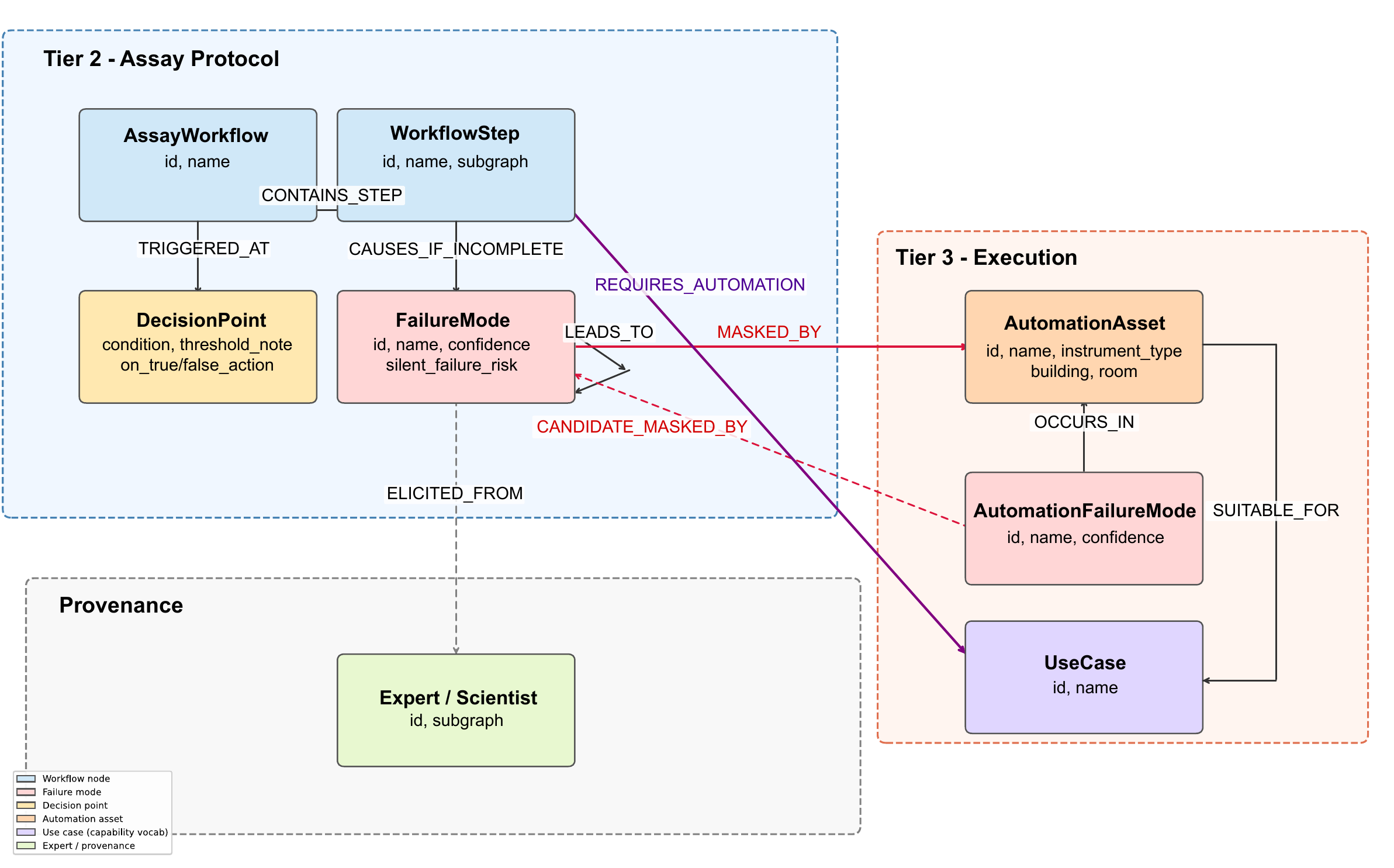}
\caption{BCP SDT node and relationship schema. The \texttt{MASKED\_BY} edge (dashed, Tier~2~$\rightarrow$~Tier~3) encodes automation-induced observability loss: a FailureMode linked to the AutomationAsset responsible for concealing it from execution logs.}
\label{fig2}
\end{figure}

The node type vocabulary is shown in Fig.~\ref{fig2} and in full tabular form in Table 2 (supplemental). Each node type carries a \texttt{subgraph} property identifying its domain scope and a per-subgraph \texttt{id}; namespace separation is enforced at the ID level.

\subsection{Edge Type Semantics}
Within-tier edges (Table 3, supplemental) define workflow structure and knowledge provenance; cross-tier edges (Table 4, supplemental) define the causal and masking connections that require the federated architecture to represent.

The \texttt{MASKED\_BY} relationship encodes a condition with no structural equivalent in existing biomedical KG literature: a scientific failure mode that is invisible to the execution layer because the automation asset logs success while scientific validity is compromised. The edge direction --- from \texttt{FailureMode} toward \texttt{AutomationAsset} --- encodes this asymmetry: the automation asset is not causing the failure; it is preventing the failure from being detected.

Two relationships form the capability bridge: \texttt{REQUIRES\_AUTOMATION} links each \texttt{WorkflowStep} to an assay-agnostic \texttt{UseCase}, and \texttt{SUITABLE\_FOR} links that \texttt{UseCase} to a physical \texttt{AutomationAsset}. The 15 \texttt{UseCase} nodes (Serial Dilution, Plate Washing, Precious Reagent Dispensing, etc.) allow workflow steps from any assay modality to reach automation assets through a shared vocabulary layer.

\subsection{Three-Subgraph Federation}
The federation spans an ELISA subgraph, an LC-MS/PRM subgraph, and an automation infrastructure subgraph. Graph provenance is fully tracked: nodes generated from text extraction carry metadata distinguishing them from expert-elicited nodes, ensuring traceability back to source scientist or protocol document. All class and edge type definitions belong in the shared upper ontology by default; instantiated edges with their assay-specific confidences reside in individual subgraphs, ensuring future additions map to upper ontology classes at construction time.

\section{Elicitation Methodology}
The SKG is populated through a structured expert elicitation pipeline centered on the Protocol Intelligence Co-pilot --- a purpose-built AI interview agent that extracts tacit procedural knowledge from domain experts and converts it into a machine-readable intermediate representation for graph ingestion. The pipeline comprises three stages: interview elicitation, structured annotation, and Cypher load into Neo4j AuraDB.

\subsection{Protocol Intelligence Co-pilot}
The Co-pilot's core design principle is that the scientist is a co-author of the knowledge object, not a subject of interrogation. After each substantive exchange, the agent updates the structured representation and displays the changed section with an explicit invitation to correct it --- surfacing misclassifications in real time while building trust in a process of explicit knowledge externalization.

\subsubsection{Session modes.} The Co-pilot operates in three modes governing which knowledge layers may be populated: \textbf{OPERATIONAL} (execution-level knowledge, for scientists who run but did not design the protocol), \textbf{DESIGN EXPERT} (full elicitation including decision model layer, for protocol designers and domain owners), and \textbf{DIRECTOR} (strategic cross-domain elicitation via a companion Director Agent, no protocol anchor).

\subsubsection{Epistemic contamination guard.} Decision model fields must never be populated from an OPERATIONAL source. When session mode is OPERATIONAL, \texttt{decision\_model} is set to \texttt{\_elicitation\_scope: ``operational\_only''} with all fields null, flagging the required follow-up: a DESIGN EXPERT session with the protocol's designer. A null field is honest --- it records that the knowledge has not yet been captured. A guessed field masquerading as design knowledge will be queried and treated as truth.

\subsubsection{Session structure.} Sessions proceed through four phases: \textbf{Orient} (anchor on the decision the assay supports, not the protocol mechanism), \textbf{Explore} (failure genealogy, conditional decision logic, procedural dependencies, tacit knowledge boundaries), \textbf{Generalize} (decision model layer, DESIGN EXPERT only), and \textbf{Close} (knowledge not surfaced by the lens structure).

\subsection{Structured Extraction Object Schema}
Each elicitation session produces a Structured Extraction Object (SEO) --- a typed JSON document that decouples elicitation from graph construction: the annotation agent operates on structured text rather than raw transcript, and the schema enforces completeness independently of which elicitation agent produced the interview.

The SEO comprises six independently grounded content layers (detailed in Appendix A); three session-mode gates govern which layers may be populated: Layer 1 (Protocol, all modes), Layer 2 (Decision Model, DESIGN EXPERT and DIRECTOR only), and Layer 3 (Strategic, DIRECTOR only: cross-domain knowledge, group capability gaps, future assay class design questions).

Every \texttt{FailureMode} and \texttt{DecisionPoint} node carries three mandatory per-node fields --- \texttt{confidence}, \texttt{confidence\_method}, and \texttt{source\_scientist} --- as specified in Layer 2 (Appendix A). Session-level provenance is recorded separately in the \texttt{twin\_metadata} block (Layer 6), anchoring each claim to its source expert, session mode, and calibration status. Fields not yet elicited remain null.

\subsection{Expert-Assigned Confidence Scoring}
Every \texttt{FailureMode} and \texttt{DecisionPoint} node in the deployed KG carries a \texttt{confidence} property in [0.60, 1.00]. Two methods are in use:
\begin{itemize}
    \item \textbf{Linguistic approximation} (primary method, both subgraphs): confidence is assigned from the expert's language during elicitation. Declarative language (``always,'' ``definitely,'' ``every time'') maps to 0.85--0.92.
    \item \textbf{SHELF elicitation~\cite{ref13}:} for failure modes where \texttt{silent\_failure\_risk: true} or \texttt{is\_critical\_path: true}, the Co-pilot elicits a three-point frequency estimate (\texttt{frequency\_min}, \texttt{frequency\_best}, \texttt{frequency\_max}) with \texttt{confidence\_method: ``SHELF\_elicited''}. SHELF and linguistic approximation populate distinct property fields: SHELF produces a frequency distribution; the scalar \texttt{confidence} used throughout \S 5 is always linguistically approximated. SHELF elicitation was applied to four silent failure mode candidates in the current deployment; formal Cooke weighting is deferred as future work.
\end{itemize}

\subsection{Graph Generation and Transcript Fidelity}
The Annotator Agent translates validated SEOs into deterministic, MERGE-compatible Cypher for idempotent ingestion into Neo4j AuraDB. A pre-processing mode extracts a baseline protocol skeleton from SOP documents; automatically extracted properties are tagged \texttt{[SCHEMA\_DEFAULT]}, while expert-validated additions are tagged \texttt{[INTERVIEW\_CONFIRMED]}. Cross-subgraph connections (e.g., \texttt{MASKED\_BY}) are flagged as PENDING CONVERGENCE and manually authored during a cross-domain validation phase, ensuring independent subgraph integrity before federation.

When elicitation uses spoken input, general-purpose ASR models introduce systematic errors by substituting phonetic approximations for scientific jargon --- errors that confidence scoring then amplifies. A dedicated contextual agent reviews raw transcripts against assay-specific vocabulary before the annotation phase, logging all corrections to preserve pipeline traceability. Across both deployed subgraphs, this step identified and corrected 288 such errors (3.5 per 1,000 characters).

\section{Evaluation: Three Subgraphs and Cross-Domain Queries}
We evaluate the federated SDT along three axes: (A) individual subgraph querying, demonstrating query types unavailable from protocol documents or sensor streams; (B) cross-subgraph federation, requiring traversal across the ELISA--Automation boundary; and (C) comparative analysis across subgraphs, showing that inter-domain variation in knowledge coverage is itself a retrievable finding.

All queries were executed against the live AuraDB deployment following full load of all three subgraphs and the upper ontology governance layer (statistics in Table 5, supplemental). All confidence scores were assigned using linguistic approximation from expert elicitation transcripts, as documented in the corresponding \texttt{CalibrationRecord} node.

\subsection{Demonstrated Query Capabilities}
Seven queries spanning six capability classes were evaluated against the live AuraDB deployment. Q1 and Q5 apply the same ranked-retrieval query to the ELISA and LC-MS subgraphs respectively; Q4 comprises two sub-queries (Q4a, Q4b) targeting distinct epistemic gap types. Full Cypher traversal queries and tabular results are provided in the Supplemental Material.

\subsubsection{Ranked Failure Mode Retrieval with Automation Visibility (Q1 \& Q5):}
The graph returns confidence-ranked failure mode catalogs for both assay subgraphs, explicitly flagging risks classified as ``SILENT'' (undetectable by connected automation). In the ELISA subgraph, the highest-confidence SILENT failure mode is Washer Carryover (confidence = 0.90), linked to the EL406 Plate Washer --- the second-ranked failure mode by expert confidence is simultaneously invisible to automated quality control. In the LC-MS subgraph, the highest-confidence failure mode is Recombinant/Endogenous Mismatch (0.90); full ranked results are in Table 6 (supplemental).

\subsubsection{Machine-Actionable Decision Logic (Q2):}
Traversals at critical workflow steps retrieve structured conditional logic including numeric thresholds, branching actions, and escalation triggers encoded from expert judgment. For the Plate Readout step in ELISA, six decision points are returned --- each encoding a typed condition, a numeric threshold, and explicit pass, fail, and escalation actions.

\subsubsection{Causal Cascade Prediction (Q3):}
\begin{figure}[htbp]
\centering
\includegraphics[width=\textwidth]{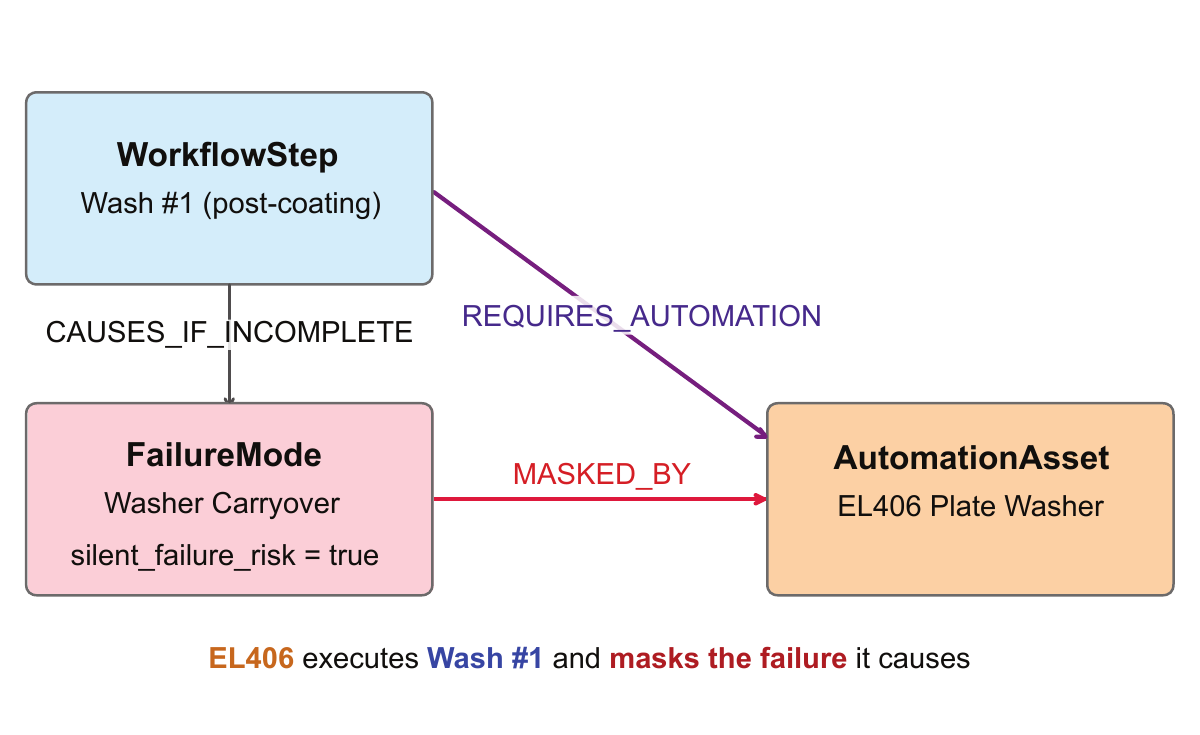}
\caption{EL406 Plate Washer self-masking loop (Q3/Q6). The automated plate washer simultaneously causes Washer Carryover (\texttt{CAUSES\_IF\_INCOMPLETE}) and prevents its detection (\texttt{MASKED\_BY}), creating an observability gap invisible to automation execution logs.}
\label{fig3}
\end{figure}

The graph reconstructs multi-step failure cascades tracing upstream procedural errors to terminal scientific invalidity. At maximum cascade depth 2 (Fig.~\ref{fig3}): Washer Carryover $\rightarrow$ High Background / Nonspecific Signal $\rightarrow$ Standard Curve Failure. The causal chain is reconstructable only through graph traversal; neither the equipment log nor the intermediate anomaly individually signals the root cause.

\subsubsection{Epistemic Self-Audit --- Coverage Gaps and Knowledge Boundaries (Q4):}
Q4 interrogates the graph's own knowledge coverage --- a capability structurally unavailable from any static document or sensor stream.

Q4a identifies ELISA workflow steps with no documented failure mode. Three steps are returned: Stop Reaction and Sample Dilution Strategy are genuine elicitation gaps; Plate Readout is structurally distinct --- it carries six decision points encoding conditional logic, but no \texttt{CAUSES\_IF\_INCOMPLETE} edge, because at readout the scientist evaluates consequences rather than executing a procedure that can fail. The distinction between a step that \emph{executes} and one that \emph{evaluates} is itself a retrievable graph finding.

Q4b identifies LC-MS failure modes at the confidence floor ($\leq 0.60$), encoding a scientist scope limitation rather than a knowledge gap. Three failure modes carry 0.60, assigned because these fell outside the elicited scientist's direct operational experience --- a structured epistemic signal, not missing data.

\subsubsection{Cross-Subgraph Federation Traversals: The EL406 Self-Masking Loop (Q6):}
The federation links scientific failure modes to the automation assets that conceal them. Both returned rows (Table 7, supplemental) resolve to the EL406 Plate Washer, which simultaneously \emph{causes} Washer Carryover and \emph{masks} it from detection (Fig.~\ref{fig3}) --- a self-masking loop with no representation in any individual data source.

\subsubsection{Cross-Assay Capability Bridge: Instrument Sharing Query (Q7):}
Querying the 31 \texttt{REQUIRES\_AUTOMATION} edges spanning both subgraphs identifies automation assets shared by ELISA and LC-MS --- candidates for cross-assay consolidation. The query returns 22 instruments across three overlap tiers (Table 8, supplemental). The 15 assay-agnostic \texttt{UseCase} nodes ensure any modality pair can be queried for overlap without modification.

\subsection{Comparative Analysis: Knowledge Coverage Across Subgraphs}
\begin{figure}[htbp]
\centering
\includegraphics[width=\textwidth]{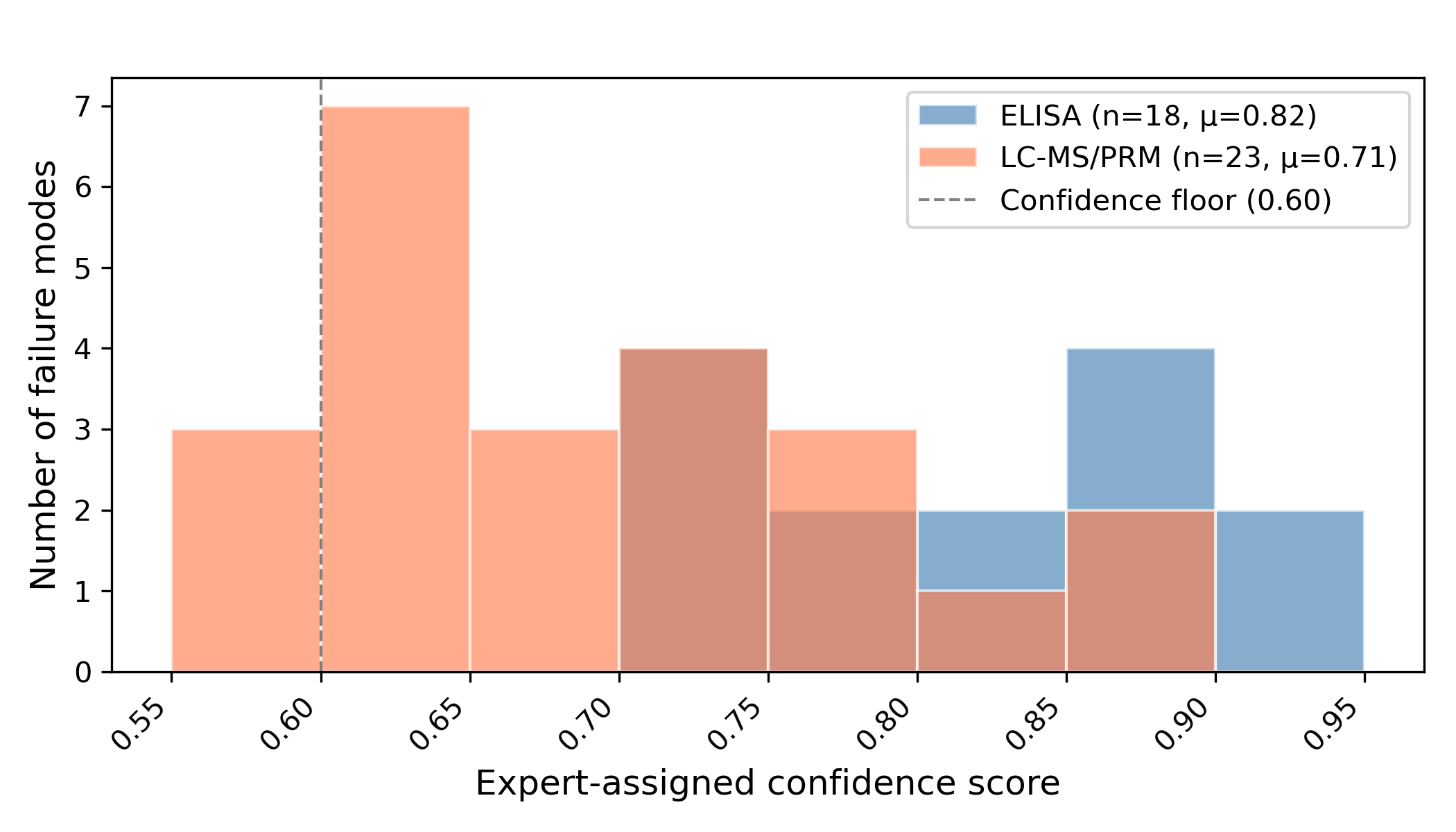}
\caption{Expert-assigned confidence score distribution by subgraph. ELISA (n=18, $\mu=0.82$) clusters toward high confidence; LC-MS/PRM (n=23, $\mu=0.71$) shows broader spread reflecting greater tacit knowledge uncertainty. Dashed line: confidence floor (0.60).}
\label{fig4}
\end{figure}

The two completed assay subgraphs share construction methodology but exhibit structurally different knowledge profiles (Fig.~\ref{fig4}, Table 9, supplemental). ELISA (n = 18, $\mu$ = 0.82) clusters toward high confidence; LC-MS/PRM (n = 23, $\mu$ = 0.71) shows broader spread reflecting greater tacit knowledge uncertainty. The structural difference in failure mode count reflects genuine domain complexity: LC-MS/PRM involves more instrument-dependent failure modes and more sample preparation chemistry steps.

A notable silent failure analogue appears in the LC-MS subgraph. Sample Evaporation / Well Edge Effect (FM-LCMS-022, confidence = 0.65) was characterized by the elicited scientist as: ``I don't think that plays a big part --- the IS corrects for all of that.'' However, SIL-IS ratio normalization mathematically masks a concentration error from evaporation rather than correcting it --- if evaporation is uniform, the IS ratio appears normal while absolute concentration drifts. The graph holds both truths simultaneously: the scientist's operational confidence (0.65) and the annotator's \texttt{silent\_failure\_risk: true}.

\subsection{Pipeline Validation: Annotator Consistency and Transcript Fidelity}

\subsubsection{Within-agent consistency.} The pipeline was run three times independently on each transcript with no shared state. All structural metrics reached 1.0 with zero variance (Table 10, supplemental). A critical distinction applies: FM F1 = 1.0 across independent runs confirms that extraction is \emph{deterministic} --- governed by elicitation content rather than annotator stochasticity. Determinism is not the same as accuracy; cross-agent agreement operates at a different evidential level.

\subsubsection{Cross-agent agreement.} The automated pipeline was compared against reference annotations from manually-guided sessions (Table 11, supplemental). On the clean-session LC-MS comparison, it achieved FM F1 = 1.0, independently extracting the identical 13 FailureMode nodes. The ELISA cross-agent comparison yielded FM F1 = 0.43 --- the automated pipeline missed failure modes surfaced only through extended multi-turn probing in the reference session. Conversely, the automated pipeline generated structurally richer output, autonomously producing \texttt{MASKED\_BY} edges that the manual sessions missed; MethodAlternative recall (0.22) directly motivates the Co-pilot's chunked, multi-turn design. The strongest cross-expert signal: recombinant/endogenous mismatch was independently recovered in both ELISA and LC-MS --- different assay domains, different scientists, the same framework (\S 5.2).

\section{Discussion}

\subsection{Challenges and Honest Limitations}
\begin{itemize}
    \item \textbf{Elicitation time investment.} Each structured session requires roughly 60 minutes of expert time plus load review --- a front-loaded cost easily justified for high-stakes assays (GLP studies, clinical biomarkers).
    \item \textbf{Tacit knowledge that resists articulation.} Some expert judgment relies on pre-verbal pattern recognition. The graph encodes these boundaries via \texttt{AmbiguityFlag} nodes and \texttt{flagged\_for\_review} properties, treating unarticulated claims as investigational targets rather than certain knowledge.
    \item \textbf{Confidence approximation.} Linguistic approximation captures graded certainty but not calibrated probability. Formal cross-scientist calibration (Cooke~\cite{ref6}) is required before the graph can support autonomous, high-stakes program decisions.
    \item \textbf{Currency maintenance.} Laboratory workflows evolve; equipment upgrades or new reagent lots can invalidate existing failure modes. Mitigation relies on periodic re-elicitation and versioned MERGE-idempotent Cypher loads; automated staleness detection remains future work.
\end{itemize}

\subsection{Broader Applicability and Future Directions}
The core components --- structured elicitation lenses, the SEO intermediate schema, and the \texttt{MASKED\_BY} representation --- are domain-agnostic. Direct candidate domains include pharmaceutical manufacturing process validation (where tacit parameter interactions are high-stakes) and clinical diagnostics (where cross-laboratory variability often encodes undocumented procedural differences). As the field progresses toward agentic laboratory systems, the SDT provides the prerequisite semantic world model: a queryable representation of where workflows fail silently, where human judgment is irreplaceable, and which automation assets mask rather than detect failures.

\section{Conclusion and Future Work}
Laboratory workflows encode substantial tacit knowledge --- expert judgment about failure conditions, decision branching logic, and contextual dependencies --- that remains inaccessible to protocol documents, sensor streams, and existing biomedical ontologies. We have presented a federated Semantic Knowledge Graph and a reproducible, structured expert elicitation methodology to capture this knowledge across three workflow domains. The deployed federation produces capabilities unachievable in isolated systems: machine-actionable decision logic, epistemic self-auditing, and pipeline extraction that is structurally deterministic within-agent (FM F1 = 1.0) with cross-agent agreement bounded by elicitation depth. Most critically, this architecture introduces the \texttt{MASKED\_BY} relationship --- formalizing a class of risk previously invisible to laboratory informatics, where execution logs report success while scientific validity is actively compromised.

Future work proceeds in three directions:
\begin{itemize}
    \item \textbf{Multi-assay expansion.} The methodology is being deployed across additional assay formats (MSD, TR-FRET, cell-based assays) with strict upper ontology alignment; expansion beyond BCP to other Genentech departments is underway.
    \item \textbf{Agent-runtime querying.} Transitioning the SKG to a dynamic reasoning substrate, enabling AI agents to query failure modes, interpret anomalous readouts, and identify masking risks in real time during experimental planning and execution.
    \item \textbf{Formal confidence calibration.} Upgrading linguistic approximation scores to calibrated probability distributions via Cooke's classical model~\cite{ref6}, prioritizing nodes with \texttt{silent\_failure\_risk: true} to maximize calibration yield while minimizing additional interview burden.
\end{itemize}

\begin{credits}
\subsubsection{\ackname}
The authors thank the domain scientists at the Biochemical and Cellular Pharmacology Department, Genentech, Inc., who contributed expert elicitation sessions. We also thank Arindam Sett, Kelly Loyet, Heather Jutila, Asif Jan, Zoe Piran, and Shirley Ng for critical reading and feedback, and Corey Bakalarski (Allotrope Foundation) for review of the ontology alignment sections.

\subsubsection*{Supplemental Material Statement:}
The supplemental document includes architecture reference tables (Tables 1--11), complete Cypher queries for all seven query types (Q1--Q7), the abridged SEO schema (Appendix A), and the upper ontology governance specification (Appendix B), submitted via EasyChair and available through the corresponding author's institutional repository upon acceptance. Co-pilot system prompts, agent prompts, elicitation lens specifications, and Cypher load files are subject to a pending patent application and are available from the corresponding author for review purposes.

\subsubsection{\discintname}
The authors are employees of Genentech, Inc., a company that sells and manufactures medicines.
\end{credits}

\section*{Declaration of Use of Generative AI}
Generative AI tools were used in this work in two capacities. First, as core research instruments: the Protocol Intelligence Co-pilot, Annotator Agent, and domain correction preprocessing agent described in Section 4 are implemented as large language model-based agents using Claude (Anthropic) and Gemini (Google); their design, application, and outputs are described in the methodology and constitute the primary contribution of this paper. Second, large language model tools --- specifically ChatGPT (OpenAI), Claude (Anthropic), and Gemini (Google) --- were used to support manuscript preparation, including deep research, literature review, structural editing, reference verification, and consistency checking. All scientific claims, experimental results, and interpretations are the sole responsibility of the human authors, who reviewed and verified all AI-assisted content.


\includepdf[pages=-]{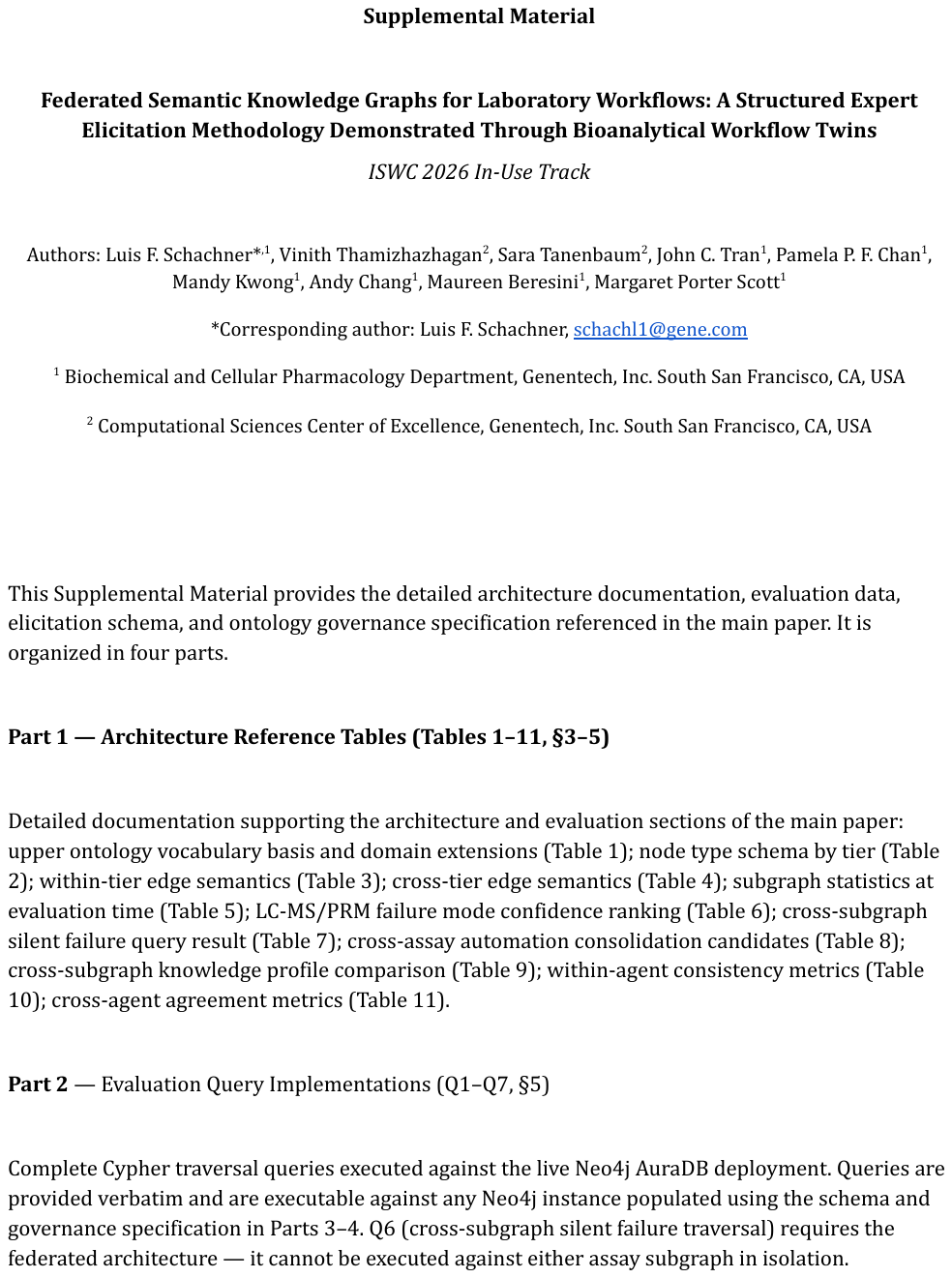}
\includepdf[pages=-]{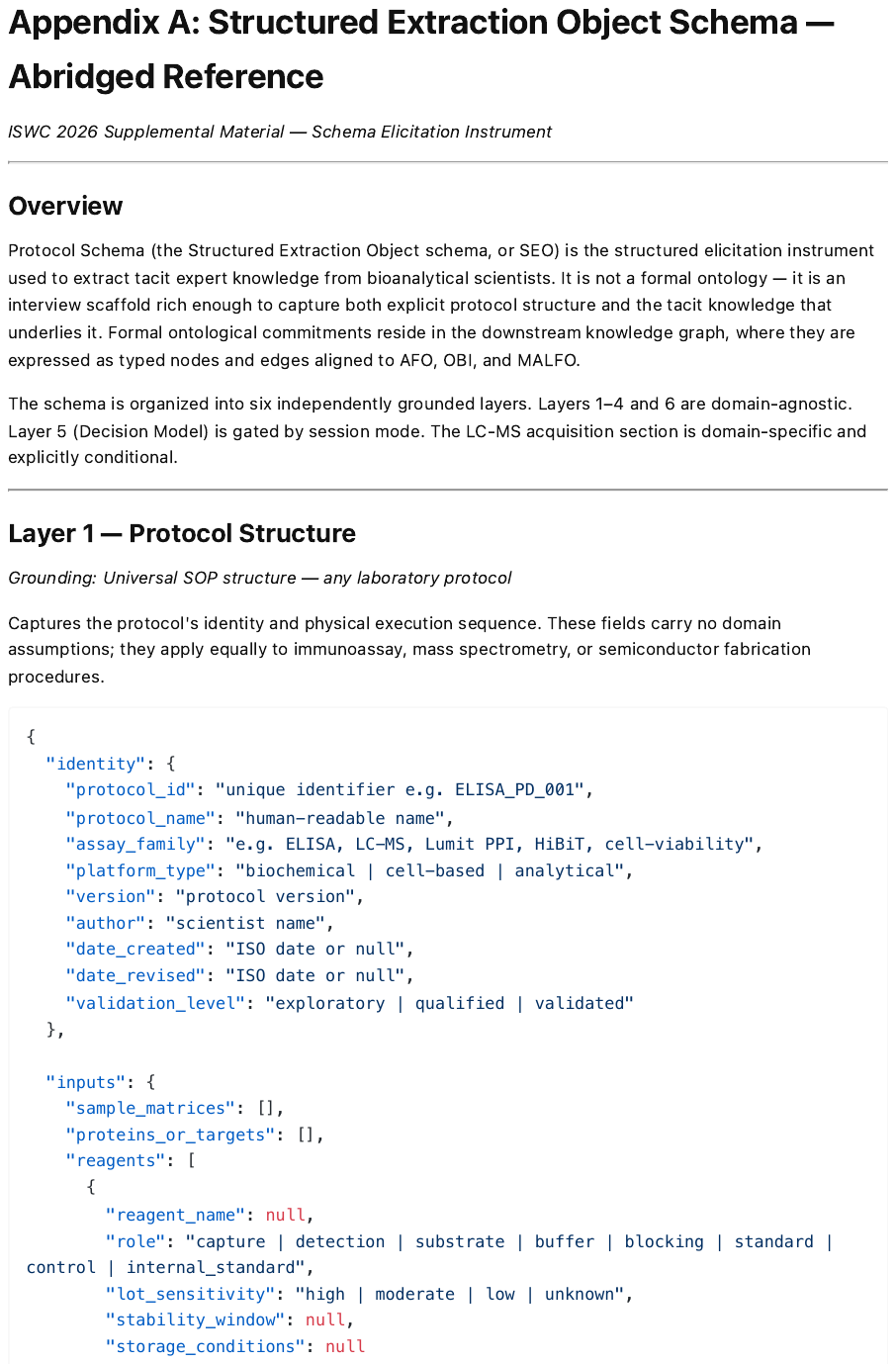}
\includepdf[pages=-]{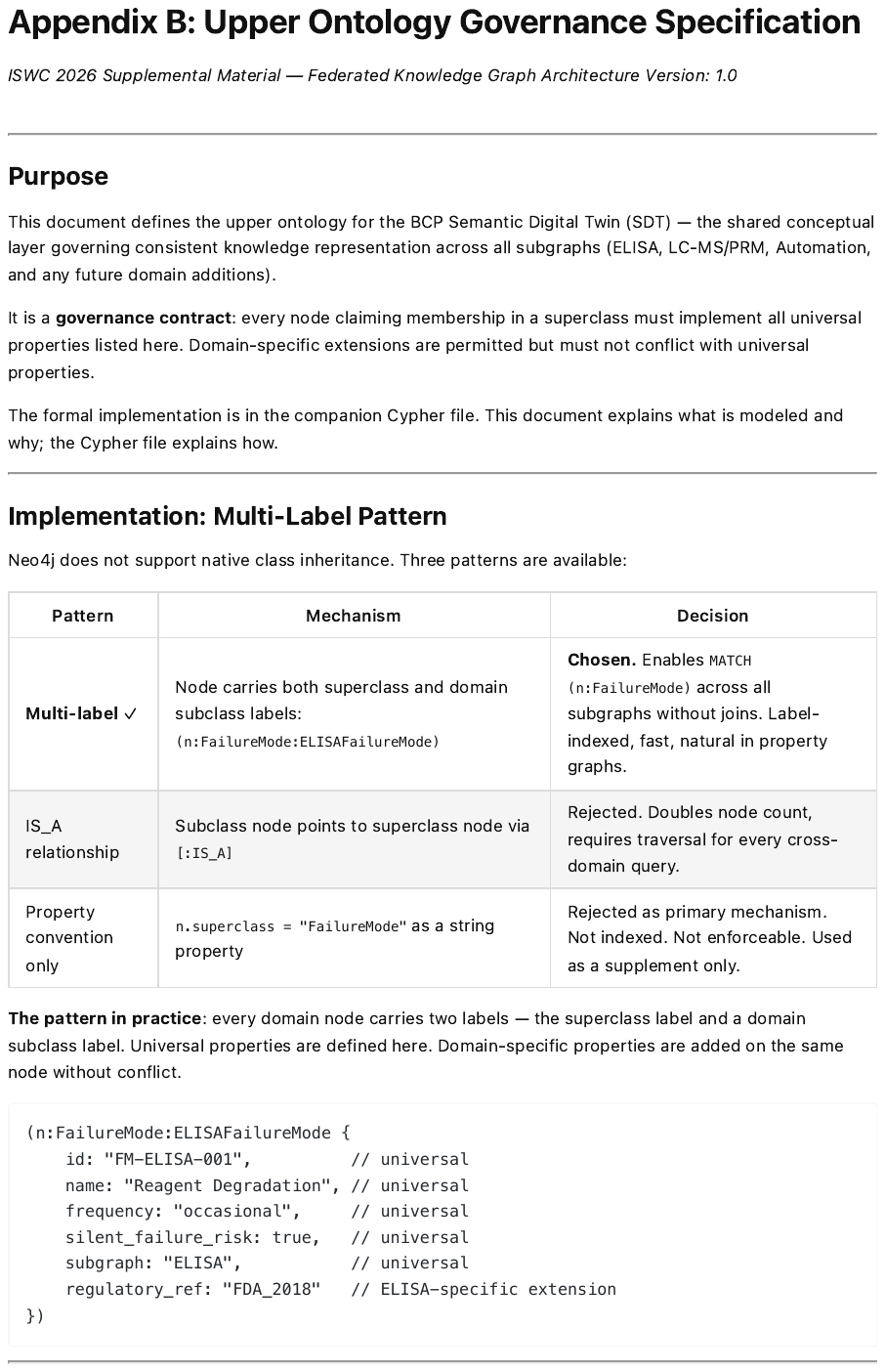}

\end{document}